\documentclass[usenatbib,usegraphicx,useAMS]{mn2e}
\usepackage{amsmath}
\usepackage{amssymb}

\newcommand{\mass}{M$_{\odot}$}
\newcommand{\persec}{s$^{-1}$}

\title{Detecting Stars at the Galactic Centre via Synchrotron Emission}

\author[Ginsburg, Wang, Loeb, Cohen]{Idan Ginsburg\thanks{E-mail:
iginsburg@cfa.harvard.edu}, Xiawei Wang\thanks{E-mail:xiawei.wang@cfa.harvard.edu}, Abraham Loeb\thanks{E-mail:aloeb@cfa.harvard.edu}
\& Ofer Cohen\thanks{E-mail:ocohen@cfa.harvard.edu} \
\\Astronomy Department, Harvard University, 60 Garden St., Cambridge, MA 02138, USA\\}

\begin{document}
\maketitle

\begin{abstract}

Stars orbiting within 1$\arcsec$ of the supermassive black hole in the Galactic Centre, Sgr A*,  
are notoriously difficult to detect due to obscuration by gas and dust.
We show that some stars orbiting this region  
may be detectable via synchrotron emission. In such instances, a bow shock forms around the star and accelerates the electrons. 
We calculate that around the 10 GHz band (radio) and at 10$^{14}$ Hz (infrared) the luminosity of a star
orbiting the black hole is comparable to the luminosity of Sgr A*. The strength of the synchrotron 
emission depends on a number of factors including the star's orbital velocity. 
Thus, the ideal time to observe the synchrotron flux is when the star is at pericenter. 
The star S2 will be $\sim 0.015\arcsec$ from Sgr A* in 2018, and is an excellent target to test our predictions.

\end{abstract}


\section{Introduction} \label{INT}

Over 100 young massive stars inhabit the central parsec of the 
Milky Way (for a review see \citealt{Genzel:10}; \citealt{Mapelli-Gualandris}). 
The stars whose orbit lies within $\sim$ 0.04 pc from the Galactic Centre (GC) 
are known as the S-stars (e.g. \citealt{Scho:03}; \citealt{Ghez:05}). 
The orbits of 28 S-stars were determined by \citet{Gillessen:2009a}. 
19 members have semimajor axis $a \leq 1$ arcsec. Of those, 16 are B stars 
and the rest late-type stars. 
Of particular interest is the star S2 (also known as SO-2) which has been 
observed for more than one complete orbit (\citealt{Ghez:05}; \citealt{Ghez:08}; \citealt{Gillessen:2009}).
S2 orbits the supermassive black hole, Sgr A*, every 15.9 years. It is a B0-2.5 V main sequence 
star with an estimated mass of $\sim$15\mass \,\citep{Martins:08}. The second complete orbit of 
a star around Sgr A* was announced not long ago \citep{Meyer:12}. This star, S102 (also know as SO-102)
has a period of 11.5 years and is about 16 times fainter than S2. Both S2 and S102 
provide compelling evidence that Sgr A* has a mass of $\sim 4\times10^6$\mass. 

The detection of young, massive stars with orbits close to Sgr A* was surprising (eg. \citealt{Ghez:03}). 
A number of possible mechanisms have been proposed to account for the origin of the S-stars. 
\citet{Lockmann:08} argued that the dynamical interaction of two stellar disks in the central parsec
could lead to the formation of the S-stars. \citet{Griv:10} proposed that the S-stars 
were born in the disk and migrated inward. Recently, \citet{Chen:14} theorized that a few Myr ago 
the disk extended down to the innermost region around Sgr A* and Kozai-Lidov-like resonance 
resulted in the S-stars. Perhaps the simplest and arguably most likely scenario is that 
at least some of the S-stars are the result of a three-body interaction 
with Sgr A* (e.g. \citealt{Ginsburg:1}; \citealt{Ginsburg:2}; \citealt{Ginsburg:4}; \citealt{Zhang:13}). 
In this scenario, a binary star system interacts with Sgr A*, and 
tidal disruption leads to one star falling into the gravitational well of the black hole while
the companion is ejected as a hypervelocity star (HVS) \citep{Hills:88}. 
HVSs were first observed in 2005 \citep{Brown:05} and as of today some 24 
have been identified (see \citealt{Brown:14} for the list). 
\citet{Brown:15} studied 12 confirmed HVSs and found that the vast majority
are consistent with having a GC origin. Furthermore, observations 
indicate that these HVSs are likely massive slowly pulsating B stars \citep{Ginsburg:5}
and thus consistent with the known S-stars. 

There are various candidate HVSs which are far less massive 
than B-type stars (\citealt{Palladino:14}; \citealt{Zhong:14}; \citealt{Zhang:15}). 
Given the fact that the stellar density in the central parsec is $\sim 10^6$\mass \,pc$^{-3}$\citep{Scho:09} 
a distribution of masses is expected. However, detecting stars at the GC is 
notoriously difficult due to dust extinction (e.g. \citealt{Scho:10}). In this paper 
we discuss using synchrotron radio emission to observe and possibly monitor stars at the GC. 
Of particular importance is the fact that S2 will reach periapse in 2017. 
We show that the closer S2 is to Sgr A* the more likely we are detect the 
synchrotron emission. Thus, S2 serves as an ideal test subject. 
In Section 2 we provide the physics behind the synchrotron emission. In Section 3 
we show that a star such as S2 may be observable via its synchrotron emission. 
We conclude with some discussions and future observations in Section 4.

\section{Shocks and Synchrotron Emission} \label{Shock} 

We consider stars orbiting Sgr A*. Interactions between winds from such a star and the interstellar medium (ISM)
will create two shocks, a reverse and forward shock. The forward shock propagates into the 
ambient medium which is swept up and subsequently compressed and accelerated. 
The faster wind is compressed and decelerated by the reverse shock.
We are interested in the emission from electrons accelerated by the forward shock. 
In our case the forward shock is a bow shock with Mach angle $\theta \sim {\cal{M}}^{-1}$, where $\cal{M}$ is the Mach number.
In such a scenario, roughly half the star's mass loss contributes to the shock. 
The mechanical luminosity is simply given by 
kinetic energy which depends upon the mass loss rate, $\dot{M}$, and wind speed
\begin{equation}
L_w = \frac{1}{2}\dot{M}_{\rm w}v_{\rm w}^2.
\end{equation}
For a massive star such as S2, typical values for $\dot{M}_w$ are $\sim 10^{-6}$ M$_{\odot}$ yr$^{-1}$,
$v_w \sim 1000$ km \persec\, and thus $L_w$ is around 10$^{35}$ erg s$^{-1}$.
Consequently, the total non-thermal luminosity, $L_{nt}$, is given by the fraction of electrons, $\epsilon_{nt}$, accelerated 
to produce non-thermal radiation
\begin{equation}
L_{nt} = \epsilon_{nt}L_w.
\end{equation}
We let $\epsilon_{nt}$ be 5\% although simulations show that this value 
is uncertain \citep{Guo:14}.
The energy density of the amplified magnetic field is given by $U_B = B^2/8\pi$. 
Therefore, assuming equipartition of energy,
$U_B = \xi_Bn_skT_s$ leads to
\begin{equation}
B = (8\pi \xi_B n_s kT_s)^{1/2}
\end{equation}
where $\xi_B$ is the fraction of thermal energy in the magnetic field, $n_s$ is 
the post-shock number density, and $kT_s$ the temperature of the post-shock medium. 
In the strong shock limit for an adiabatic index of 5/3,
the post-shock number density $n_s$ is $\sim 4$ times the number density of the ambient ISM. 
Although the value of $\xi_B$ is highly uncertain, a value of 0.1 is reasonable \citep{Volk:05}.
The resulting values for $B$ are $\sim 10^{-2} - 10^{-3}$ G. 
The Rankine-Hugoniot jump conditions give the post shock temperature
\begin{equation}
T_s = \frac{[(\gamma_t -1){\cal{M}}^2 + 2][2\gamma_t {\cal{M}}^2 - (\gamma_t - 1)]}{(\gamma_t + 1)^2{\cal{M}}^2}T_o
\end{equation}
where $T_o$ is the upstream temperature, and $\gamma_t$ is the 
adiabatic index which is taken to be 5/3. 
At the contact discontinuity, ram pressure from the star/ISM is balanced by ram pressure from the wind. Thus we have
$\rho_*v_*^2 = \rho_w v_w^2$ where the mass flux is given as $\rho_w v_w = \frac{\dot{M}_w}{4\pi R^2}$
and this leads to the standoff radius
\begin{equation}
R_o = (\frac{\dot{M}_w v_w}{4\pi \rho_* v_*^2})^{1/2}. 
\end{equation}
$\rho_* = n_om_p$ where $m_p$ is the proton mass. 
At a distance of around 0.3 pc from Sgr A* 
$n_o \sim 10^3$ cm$^{-3}$ and the temperature of the ISM is $\sim 10^7$ K \citep{Quataert:04}.  
Assuming $v_* = 1000$ km \persec\, a typical standoff radius is approximately $10^{-3}$ pc.


We consider a broken power law distribution of electrons generated via Fermi acceleration, written as
\begin{equation}
N(\gamma)d\gamma = K_o\gamma^{-p}(1 + \frac{\gamma}{\gamma_b})^{-1} (\gamma_{min} \leq \gamma \leq \gamma_{max})
\end{equation}
where $K_o$ is the normalization factor in electron density distribution, $p$ is the electron power law
distribution index which in our calculations is $\sim 2$. 
$\gamma$ is the Lorentz factor, $\gamma_{min}$ and $\gamma_{max}$ are the minimum and maximum Lorentz factor respectively,
and $\gamma_b$ is the break Lorentz factor due to synchrotron cooling. The total power from synchrotron
emission of a single electron is given by
\begin{equation}
P = \frac{4}{9}r_o^2 \beta^2 \gamma^2 B^2
\end{equation}
where $r_o$ is the classical electron radius and $\beta \equiv v/c$ \citep{Rybicki-Lightman}.
The corresponding synchrotron cooling time is
\begin{equation}
t_{cool} = \frac{\gamma mc^2}{P}.
\end{equation}
The break Lorentz factor due to synchrotron cooling is obtained by
equating the synchrotron cooling times scale and the dynamical timescale 
\begin{equation}
t_d = \frac{R_o}{v_w}
\end{equation}
where $R_o$ is the standoff radius and $v_w$ the wind velocity. 
The peak luminosity lasts for the pericenter crossing timescale with typical values $\sim 1$ yr.
The precise shape of the peak depends on the orbital parameters, the ambient gas distribution, and the wind
mass loss rate. The characteristic timescale for the peak is the pericenter crossing time, $\sim b/v$.
$\gamma_{min}$ is set to one in our calculations. $\gamma_{max}$ is obtained by 
equating the acceleration timescale, $t_{acc} = \xi_{acc}R_Lc/v_w^2$ \citep{Blandford-Eichler}, to the dynamical or cooling timescale
where $\xi_{acc}$ is a dimensionless parameter on the order of unity.
The Larmor radius is given by $R_L = \gamma m_ec^2/eB$ where $m_e$ is the mass of the electron and $e$ is the electron charge. 
We find that $t_{acc} \sim 10^{-4}$ yr, which is far shorter than the dynamic time. 
At radio frequencies the cooling time is a few orders of magnitude greater than the dynamical time and
therefore synchrotron cooling is negligible. Thus, we assume that all 
the shocked electrons contribute to the observed synchrotron emission. 

The synchrotron flux and power are computed using the radiative transfer equation \citep{Rybicki-Lightman}
which leads to
\begin{equation}
I_\nu = \frac{j_\nu}{\alpha_\nu} (1 - e^{-\tau_\nu})
\end{equation}
where $j_\nu$ is the emission coefficient, $\alpha_\nu$ the absorption coefficient, and $\tau_\nu$ 
the optical depth (see \citealt{Wang-Loeb} for further details).

\section{Emission Around Sgr A*} \label{S2}

\begin{figure*}
\begin{center}
\includegraphics[width=\textwidth]{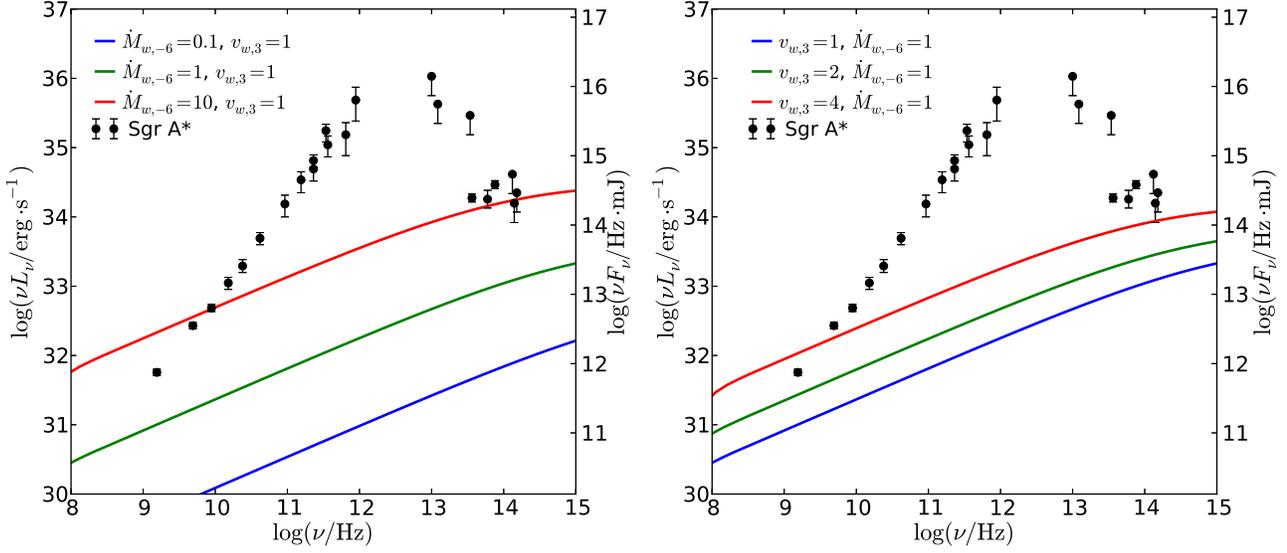}
\end{center}
\caption{
Non-thermal synchrotron power and flux compared with emission from Sgr A* 
(data for Sgr A* was obtained from \citealt{Yuan-Narayan}). The left panel shows the dependence 
of the synchrotron emission on wind mass loss rate. The star's velocity and wind velocity are both fixed at 1000 km \persec.
Mass loss, {$\dot{M}$}, was computed with values of $10^{-7}$\mass\, yr$^{-1}$ (blue line), 
$10^{-6}$\mass\, yr$^{-1}$ (green line), and $10^{-5}$\mass\, yr$^{-1}$ (red line). 
In the right panel we kept the mass loss constant at $10^{-6}$\mass\, yr$^{-1}$ and used
wind velocities of 1000 km \persec (blue line), 2000 km \persec (green line),
and 4000 \persec (red line). 
}
\label{fig1}
\end{figure*}

Figure \ref{fig1} shows the expected synchrotron flux for a star orbiting Sgr A*.
For the left panel we kept the wind velocity constant at 1000 km\persec. 
We extrapolated the particle density to have the value $\sim 10^4$ cm$^{-3}$ from \citet{Quataert:04}.
The mass loss rate for hot massive stars is poorly constrained (for a review see \citealt{Puls:08}). 
Thus, we varied the mass loss rate between $10^{-7}$\mass\, yr$^{-1}$ and  
$10^{-5}$\mass\, yr$^{-1}$ which are acceptable values for hot massive stars 
such as the S-stars \citep{Dupree}. Given that 
\begin{equation}
F_{\nu} = \frac{\nu L_{\nu}}{4\pi d^2}\frac{1}{\nu}
\end{equation}
where $d$ = 8 kpc we get a flux between $\sim 10 - 1000$ mJy
in the GHz range. 
Recently, \citet{Yusef-Zadeh:arXiv} described radio observations of over 40 massive stars
within $30\arcsec$ of the GC. 
Their values for the flux are consistent with our results.
Similar results were obtained when we kept the mass loss rate constant at 
$10^{-6}$\mass\, yr$^{-1}$ and varied the wind velocity between 1000 km \persec\, and 4000 
km \persec\, (see right panel of Figure \ref{fig1}). 

\begin{figure}
\begin{center}
\includegraphics[width=0.5\textwidth]{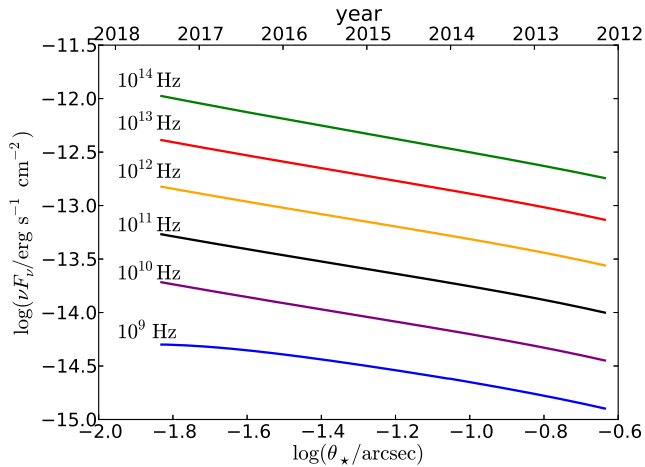}
\end{center}
\caption{
Synchrotron power (arcseconds) versus distance (erg \persec\,cm$^{-2}$) from Sgr A* for star S2. 
We let $\dot{M} = 10^{-6}$ M$_{\odot}$ yr$^{-1}$. 
At a frequency of around 1 GHz, S2 is quite bright with a luminosity of $\sim$ 10 mJy. 
The synchrotron power will be greatest in 2017, when S2 is at periapse. 
}
\label{fig2}
\end{figure}

In Figure \ref{fig2} we plot the synchrotron power versus distance from Sgr A* for S2. 
At periapse S2 has a speed of $\sim$ 5500 km \persec\, while at apoapse the speed drops 
down to $\sim 1300$ km \persec. A higher speed leads to a stronger shock. However, even 
at apoapse it may very well be possible to observe the synchrotron emission from S2. 
At 10 GHz, S2 has a flux density of $\sim$ 10 mJy. Thus, S2 is an excellent target to 
observe. It is important to note that radio wavelength photons are scattered and the image size
follows a $\lambda^2$ dependence (e.g. \citealt{Bower:06}; \citealt{Fish:14}) of
$\sim 1$ mas ($\lambda$ /cm)$^2$. Therefore, in order to resolve S2
at the 0.01\arcsec \,level, we need to observe at a frequency of 10 GHz or greater. 
Sgr A* has been observed in the radio for decades. \citet{Kellermann:77} 
used very long baseline interferometry (VLBI) to observe Sgr A* at 7.8 GHz, and they  
detected a nearby secondary transient source which has not been explained, and 
could in fact be S2 or another star.
Furthermore, \citet{Herrnstein:04} monitored the flux density of Sgr A* using the Very Large Array. 
Their results are consistent with our calculation. 
\citet{Macquart-Bower} looked at the long-term variability of Sgr A* 
and detected a flux of $\sim 10^{31}$ erg \persec \,around the 1987 pericenter passage of S2. 
This indicates that the mass loss rate of S2 is $\sim 10^{-6}$ \mass\, yr$^{-1}$. 
However, these data points are sparse and hence do not provide a tight constraint on the 
peak flux during the future pericenter passage.

\begin{figure*}
\begin{center}
\includegraphics[width=\textwidth]{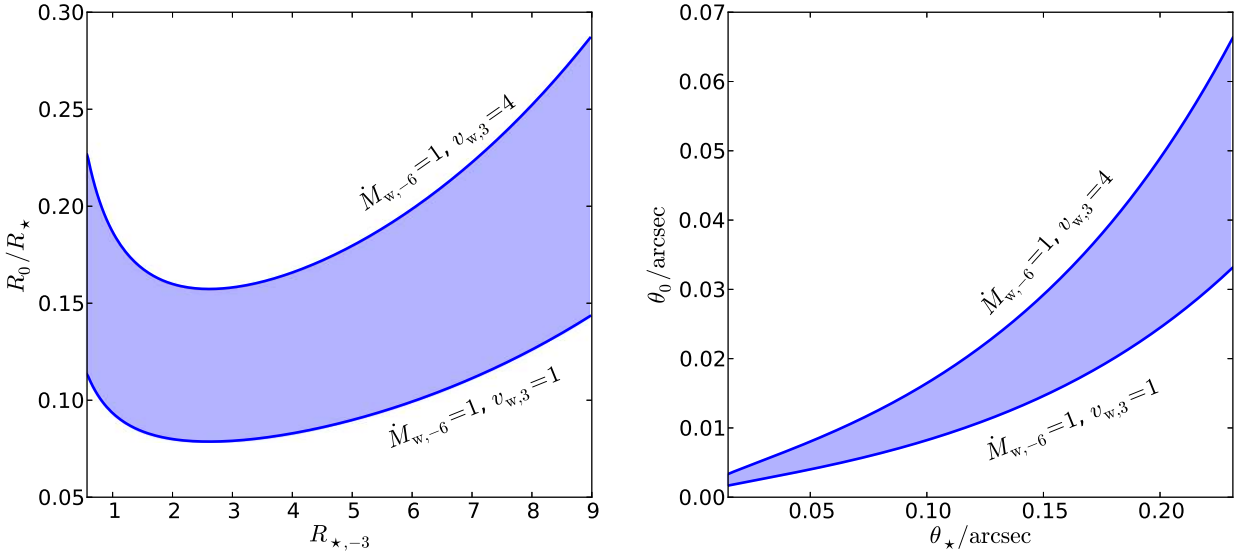}
\end{center}
\caption{
Left: The standoff radius for S2 (R$_o$) versus the orbital radius (R$_\star$ in units of $10^{-3}$ pc). 
We see that the value is always less than unity, as expected. Right: angular diameter of the standoff radius ($\theta_o$)
versus the angular diameter of the orbital radius ($\theta_\star$). 
Around pericenter it may be difficult to resolve the finite size of the standoff radius. However, S2 should still be 
detectable via synchrotron emission. 
}
\label{fig3}
\end{figure*}

While we expect the synchrotron emission to be detectable, the same is not necessarily 
true for the standoff radius. In Figure \ref{fig3} we plot the standoff radius versus orbital
radius. The standoff radius near the pericenter may not be resolvable, although at larger
distances from Sgr A* the likelihood for resolving the source increases. 
It is also worth noting that we calculated the thermal free-free emission around Sgr A*
and concluded that it is not detectable.

\section{Conclusions and Future Observations}
\label{conclusion}

The innermost stars orbiting Sgr A* (such as the S-stars) produce a bow shock. 
Although this shock will likely not be detectable via thermal emission, 
we have shown that stars such as S2 may be detectable via synchrotron emission. 
Figure \ref{fig1} shows that the contrast of synchrotron flux relative to Sgr A* is maximized around the 1.4 GHz band. 
However, due to scattering we will need to use a higher frequency, around 10 GHz, if we are to resolve our star. 
If a star such as S2 emits strong synchrotron emission, the combined signal
from the star and Sgr A* may exceed the quiescent radio emission from the black hole. 
As was the case with G2 (see next paragraph) it is not clear that any additional synchrotron emission will be observable. 
Arguably, it is best to resolve the synchrotron emission from any star orbiting Sgr A*, such as S2. 
At apoapse S2 is $\sim 0.23\arcsec$ from Sgr A* while at periapse
it is only $\sim 0.015\arcsec$ away. Thus, to detect a star such as S2 
requires both good sensitivity and resolution. 
VLBI can provide submilliarcsec observations of Sgr A* (e.g. \citealt{Lu:11}) and thus the required precision. 
S2 is an ideal test case since the star will reach periapse in 2017 or early 2018. 

Recently, gas cloud G2 was observed orbiting Sgr A*, and was approximately 3100 Schwarzschild radii 
from the black hole at pericenter. It was predicted that the bow shock 
from this encounter would displace the quiescent radio emission of 
Sgr A* by $\sim 33$ mas (\citealt{Narayan:12}; \citealt{Sadowski:13}).
However, observations across the spectrum showed no apparent variability 
during the periastron passage of G2 (\citealt{Bower:15}; \citealt{Valencia-S:15}).
It is unclear why a bow shock was not detected. 
One possibility is that at the center of G2 is low-mass star 
with wind velocity $\sim 100$ km \persec\, (\citealt{Crumley-Kumar}; \citealt{Scoville-Burkert}).
Synchrotron emission from such a small $v_w$ would be extremely difficult to detect. 
Star S2 is massive, and thus the winds are likely an order of magnitude larger. 
If synchrotron emission is not detected for S2 it may be  
that our value for $\epsilon_{nt}$ is too large. It 
may also be that the wind speeds from S2 are lower than expected, or that 
the number density around Sgr A* is significantly less than what is predicted. 
Arguably, the most likely reason why synchrotron emission from S2 may not be detected is 
simply that the shock is not as strong as assumed. 
Our results are valid so long as the shock is strong, but the emission 
could be much fainter if the strength of the shock is lessened. 
Since the shock will be strongest when S2 passes periapse, this 
will be the ideal time to monitor the star for synchrotron emission.
Even a null detection will help place constraints on the 
environment around Sgr A* \citep{Giannios:13}.

In addition to radio observations at around the 10 GHz band, Figure \ref{fig1} shows that 
at infrared wavelengths, in particular around $10^{14}$ Hz, Sgr A* is approximately
the same luminosity as a close-by star such as S2. 
However, S2 itself emits in the infrared, therefore we will not be able to distinguish 
between the non-thermal and thermal emission. 
We will need to compare the measured flux with that of the S2.
An instrument with enough sensitivity, such as the James Webb Space Telescope (JWST),
should be able to detect the synchrotron emission in the infrared. 
JWST is scheduled to launch 
in October of 2018. While S2 will have passed pericenter, if our predictions are correct 
the synchrotron emission should still be detectable. 
Furthermore, it may be possible to measure the increase in total infrared emission from S2
between apocenter and pericenter owing to the increased synchrotron emission.
While S2 is the litmus test for detecting 
stars at the Galactic Centre via synchrotron emission, we hope that ultimately we can use 
this technique to detect and monitor stars that have thus far not been observable. 

\section*{Acknowledgments}

We thank Michael D. Johnson, Atish Kamble, Robert Kimberk, Mark Reid, Lorenzo Sironi, and Ryo Yamazaki for helpful comments. 
This work was supported in part by Harvard University and NSF grant AST-1312034.

\bibliographystyle{mn2e.bst}
\bibliography{HVP.bib}
\bsp

\end{document}